# Resummed thermodynamic perturbation theory for bond cooperativity in associating fluids with small bond angles: Effects of steric hindrance and ring formation

Bennett D. Marshall[1], Amin Haghmoradi and Walter G. Chapman

*Department of Chemical and Biomolecular Engineering, Rice University, 6100 S. Main, Houston, Tx 77005*

## Abstract

In this paper we develop a thermodynamic perturbation theory for two site associating fluids which exhibit bond cooperativity (system energy is non – pairwise additive). We include both steric hindrance and ring formation such that the equation of state is bond angle dependent. Here the bond angle is the angle separating the centers of the two association sites. As a test, new Monte Carlo simulations are performed, and the theory is found to accurately predict the internal energy as well as the distribution of associated clusters as a function of bond angle.

---

[1] Author to whom correspondence should be addressed
   Email: bennettd1980@gmail.com

# 1. Introduction

Wertheim's thermodynamic perturbation theory[1-5] (TPT) provides an accurate and simple method to predict the properties of associating fluids. Here an associating fluid is meant to describe molecules which have short ranged directional interactions which saturate (e.g. the hydrogen bond). In its simplest and most widely used form[6], first order perturbation theory (TPT1), each association site is treated independently meaning there is no information of bond angle (angle between center of association sites) in the theory. Recently Marshall and Chapman[7] showed that this approximation is valid for large bond angles; however, for bond angles $< 90°$ additional information needs to be included in the theory. For these small bond angles steric hindrance between association sites, ring formation, and (for very small bond angles) double bonding must be accounted for. Marshall and Chapman[7, 8] included each of these features into a new TPT which was explicitly dependent on bond angle and shown to be highly accurate for the prediction of the distribution of associated clusters as well as the thermodynamics.

One of the fundamental assumptions in the development of TPT is that the system energy is given as the pairwise additive sum of interactions between different molecules. There is no hydrogen bond cooperativity (HBC). Of course there are many situations in nature where HBC does occur. Both hydrogen fluoride[9] and alcohols[10] exhibit strong HBC. Also, HBC has been shown to stabilize peptide hydrogen bonds[11]. To extend TPT to include HBC, Marshall and Chapman[12] recently developed a new TPT which treated bond cooperativity as a perturbation. The bond cooperativity perturbation was treated in infinite order allowing for a summation over all chain graphs. The resulting theory was surprisingly simple and shown to be highly accurate in comparison to simulation results.

Both the incorporation of bond angle dependence and HBC represent significant advances of TPT. However, the case of HBC in the presence of ring formation and steric hindrance has not been addressed. For example, quantum calculations have shown that HBC in hydrogen fluoride stabilizes ring formation.[9] Now the question must be asked, "Can we include steric effects and ring formation in a TPT for fluids which exhibit HBC?" This will be the subject of this paper. We will consider a two site associating fluid with bond angles such that both steric hindrance and ring formation must be accounted for. The extension of the contribution due to ring formation to the HBC case is trivial; however, the development of the contribution for association into linear chains which exhibit bond angle dependence and HBC is much more challenging.

In section 2 we develop the new theory. It will be shown that the inclusion of HBC adds little complexity to our previous theory[7] for bond angle dependence in two site associating fluids. In section 3 we compare the theory to new Monte Carlo simulation results. It is shown that HBC has a significant effect on the types of associated clusters which are formed. The theory is shown to be accurate in comparison to simulation results. Finally in section 4 we give conclusions.

## 2. Theory

In this section we develop the resummed thermodynamic perturbation theory for HBC in two site associating fluids with a single type *A* and type *B* association site. We restrict association such that there are *AB* attractions but no *AA* or *BB* attractions. Unlike our previous paper on HBC (we will refer to this paper as **I**) which assumed large bond angles $\alpha_{AB}$ (the angle between the centers of the association sites), here we will allow a wide range of bond angles such that steric hindrance and ring formation need to be accounted for (similar to our previous paper for bond angle dependence[7] which we will refer to this paper as **II**). As in **I**, we follow Sear and Jackson and consider a fluid composed of $N$ hard spheres of diameter $d$ with two association sites $A$ and $B$ with a total energy composed of pairwise and triplet contributions[13]

$$U(1...N) = \frac{1}{2}\sum_{i,j}(\phi_{HS}(r_{ij}) + \phi_{as}^{(2)}(ij)) + \frac{1}{6}\sum_{i,j,k}\phi_{as}^{(3)}(ijk) \tag{1}$$

where $(1) = \{\vec{r}_1, \Omega_1\}$ represents the position $\vec{r}_1$ and orientation $\Omega_1$ of sphere 1 and $\phi_{HS}$ is the hard sphere reference potential. The terms $\phi_{as}^{(2)}(ij)$ and $\phi_{as}^{(3)}(ijk)$ are the pairwise and triplet association contributions and are given by[13]

$$\phi_{as}^{(2)}(ij) = -\varepsilon^{(1)}(O_{AB}(ij) + O_{BA}(ij))$$

$$\begin{aligned}\phi_{as}^{(3)}(ijk) = -(\varepsilon^{(2)} - \varepsilon^{(1)})(&O_{AB}(ij)O_{BA}(ik) + O_{BA}(ij)O_{AB}(ik) \\ &+ O_{AB}(ji)O_{BA}(jk) + O_{BA}(ji)O_{AB}(jk) \\ &+ O_{AB}(ki)O_{BA}(kj) + O_{BA}(ki)O_{AB}(kj))\end{aligned} \tag{2}$$

Where $O_{AB}(ij)$ is the association site overlap function which, in this paper, we obtain using conical square well association sites[14-16]

$$O_{AB}(ij) = \begin{cases} 1 & r_{12} \leq r_c \text{ and } \theta_A \leq \theta_c \text{ and } \theta_B \leq \theta_c \\ 0 & \text{otherwise} \end{cases} \tag{3}$$

which states that if spheres $i$ and $j$ are within a distance $r_c$ of each other and each sphere is oriented such that the angles between the site orientation vectors and the vector connecting the two spheres, $\theta_A$ for sphere $i$ and $\theta_B$ for sphere $j$, are both less than the critical angle $\theta_c$ the two sites are considered bonded. See Fig. 1 for an illustration. The triplet contribution $\phi_{as}^{(3)}$ serves to add a correction $-(\varepsilon^{(2)} - \varepsilon^{(1)})$ for each sphere bonded twice. With this potential an associated chain of $n$ spheres will have a cluster energy $\varepsilon_{ch}^{(n)} = -\varepsilon^{(1)} - (n-2)\varepsilon^{(2)}$ and an associated ring of $n$ spheres will have an energy $\varepsilon_{ring}^{(n)} = -n\varepsilon^{(2)}$.

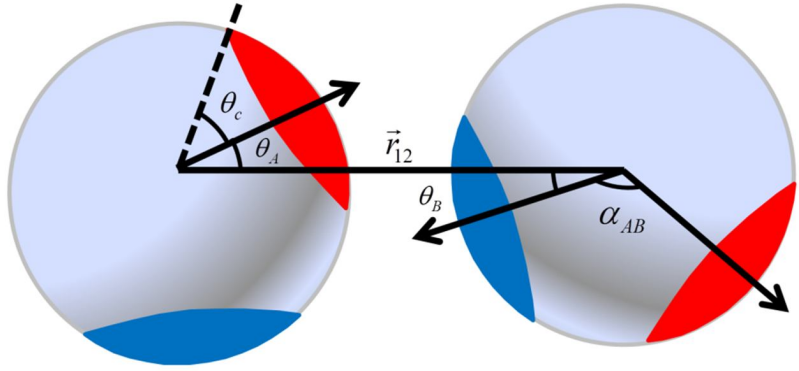

**Figure 1:** Diagram of interacting spheres with two association sites. The angular width of the association sites is determined by the critical angle $\theta_c$ and the centers of the sites are separated by the angle $\alpha_{AB}$

In the application of the theory it will be necessary to partition these cluster energies among the various bonds in the cluster. For the case of a ring, the obvious way to partition $\varepsilon_{ring}^{(n)}$ is to give each bond an energy of $-\varepsilon^{(2)}$. For the case of a chain we follow the same convention as in **I** and give the first bond in the chain an energy $-\varepsilon^{(1)}$ and each subsequent bond an energy $-\varepsilon^{(2)}$. Figure 2 gives the resulting effective bond energy distribution for associated clusters consisting of 4 monomers.

In Wertheim's multi-density formalism for two site associating fluids the Helmholtz free energy is given by[5]

$$\frac{A - A_{HS}}{k_B TV} = \rho \ln\frac{\rho_o}{\rho} - \sigma_A - \sigma_B + \frac{\sigma_A \sigma_B}{\rho_o} + \rho - \frac{\Delta c^{(o)}}{V} \tag{4}$$

Here $T$ is the temperature, $k_B$ is the Boltzmann constant, $\rho$ is the total density, $\rho_o$ is the monomer density and $\sigma_A = \rho_A + \rho_o$ where $\rho_A$ is the density of molecules bonded at only site $A$.

There is a similar relation for $\sigma_B$. The term $V$ is the system volume and $A_{HS}$ is the free energy of the hard sphere reference system. Finally, $\Delta c^{(o)}$ is the associative contribution to the fundamental graph sum which encodes all association interactions.

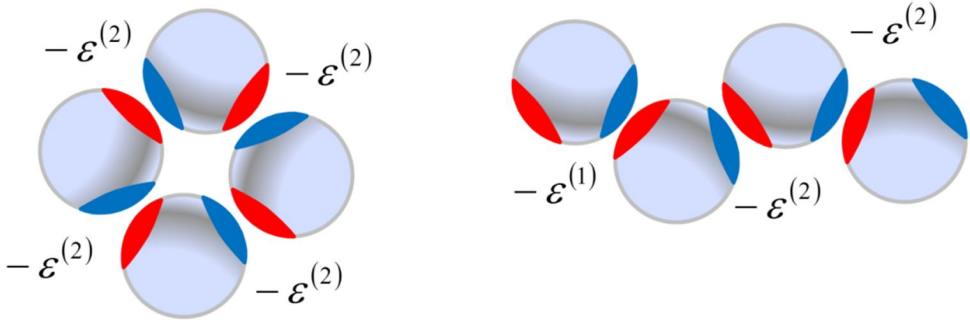

**Figure 2:** Diagram of effective bond energy distributions in associated clusters of 4 monomers

To evaluate $\Delta c^{(o)}$ we will consider molecules with small to large bond angles, but we restrict the bond angles to $\alpha_{AB} > 2\theta_c$ such that double bonding[7] between molecules cannot occur. We will also assume each association site is singly bondable. For this case

$$\Delta c^{(o)} = \Delta c_{ch}^{(o)} + \Delta c_{ring}^{(o)} \tag{5}$$

where $\Delta c_{ch}^{(o)}$ is the contribution due to the formation of chains of association bonds and $\Delta c_{ring}^{(o)}$ accounts for rings of association bonds. Each of these contributions will be strongly dependent on $\alpha_{AB}$.

The contribution due to ring formation is a very simple extension of the results of **II**. We write $\Delta c_{ring}^{(o)}$ as a sum over contributions for rings of size $n$

$$\Delta c_{ring}^{(o)} = \sum_{n=3}^{\infty} \Delta c_{ring}^{(n)} \tag{6}$$

Each ring size contribution is given as

$$\frac{\Delta c_{ring}^{(n)}}{V} = \frac{\left(f_{AB}^{(2)} \rho_o \hat{g}_{HS} K\right)^n}{nd^3} \Gamma^{(n)} \qquad (7)$$

where $f_{AB}^{(2)} = \exp(\varepsilon^{(2)}/k_B T) - 1$ and $K = \pi(1-\cos\theta_c)^2(r_c - d)$. The integral $\Gamma^{(n)}$ is proportional to the partition function of an isolated ring which is independent of density and temperature. $\Gamma^{(n)}$ is strongly bond angle dependent, with numerical results given in **II**. As can be seen in Fig. 3, for each ring size there is an optimum bond angle which maximizes the probability of ring formation. Finally $\hat{g}_{HS}$ is given by

$$\hat{g}_{HS} = g_{HS}(d) \frac{2^p}{(r_c/d + 1)^p} \qquad (8)$$

The term $g_{HS}(d)$ is the contact value of the hard sphere reference pair correlation function and $p$ is a density dependent polynomial $p = 17.87\eta^2 + 2.47\eta$ where $\eta = \pi d^3 \rho / 6$ is the packing fraction. Equation (7) was obtained by assuming the following approximation of $g_{HS}(r)$ within the bonding volume[17]

$$g_{HS}(r) = g_{HS}(d)\left(\frac{d}{r}\right)^p \quad for \quad d \leq r \leq r_c \qquad (9)$$

The only difference between Eq. (7) and [Eq. (18) of **II**] is the exchange $f_{AB} \rightarrow f_{AB}^{(2)}$, which is a result of the fact that each sphere is bonded twice in conjunction with the defined HBC.

The evaluation of $\Delta c_{ch}^{(o)}$ for the current case is more challenging. To account for chains of association bonds, we must derive $\Delta c_{ch}^{(o)}$ in a resummed perturbation theory (RTPT) which accounts for the fact that association at one site can block association at the other as in **II**, as well as incorporates the effect of HBC as in **I**. Our starting place is Wertheim's RTPT solution in the absence of HBC[5]

$$\frac{\Delta c_{ch}^{(o)}}{V} = \frac{\sigma_A \sigma_B \Xi}{1 + (1-\Psi)\rho_o \Xi} \qquad (10)$$

For our current association pair potential $\phi_{as}^{(2)}(ij)$, $\Psi$ is the probability that if sphere 1 is bonded to site $A$ on sphere 2, and sphere 3 is bonded to site $B$ on sphere 2, that there is no overlap between spheres 1 and 3. This quantity was calculated in **II** and is illustrated in Fig. 3. For small bond angles the probability there is no overlap is small giving $\Psi \to 0$ as $\alpha_{AB} \to 0°$, while for large bond angles the effect of steric hindrance is small giving $\Psi \to 1$. Steric hindrance begins to have a significant effect for bond angles $< 90°$. The term $\Xi$ is given by

$$\Xi = \sum_{m=1}^{\infty} \rho_o^{m-1} E_m \qquad (11)$$

Where the contributions $E_m$ account for associated clusters consisting of chains of $m$ bonds. In the absence of HBC $\varepsilon^{(1)} = \varepsilon^{(2)}$ this contribution is given by (after redefining some terms)[5]

$$E_m = \frac{1}{\Omega^m} \int \hat{E}(1 \cdots m+1) d\vec{r}_2 d\Omega_2 \cdots d\vec{r}_{m+1} d\Omega_{m+1} \qquad (12)$$

Where $\Omega = 8\pi^2$ for the non-axially symmetric case and the integrations are over the positional and orientational degrees of freedom of molecules 2 through $m+1$ in the cluster. The first few $\hat{E}(1 \cdots m+1)$ are given as

$$\hat{E}(12) = s(12) \qquad (13)$$

$$\hat{E}(123) = s(123) - s(12)e_{HS}(13)s(23)$$

$$\hat{E}(1234) = s(1234) - s(123)e_{HS}(24)s(34) - s(12)e_{HS}(13)s(234)$$
$$\quad + s(12)e_{HS}(13)s(23)e_{HS}(24)s(34)$$

The terms $s(1\cdots k) = f_{AB}^{(1)}(12)\cdots f_{AB}^{(1)}(k-1,k)g_{HS}(1\cdots k)$ give the product of all association Mayer functions and the *k* body reference system correlation functions for a chain consisting of *k* spheres. The Mayer functions $f_{AB}^{(j)}(12)$ are defined as

$$f_{AB}^{(j)}(12) = \left(\exp(\varepsilon^{(j)}/k_B T) - 1\right)O_{AB}(12) = f_{AB}^{(j)}O_{AB}(12) \tag{14}$$

The terms $e_{HS}(12)$ in Eq. (13) are the reference system *e* bonds which vanish when there is hard core overlap, and are unity otherwise. The general method to determine $\hat{E}(1\cdots k)$ is to take $s(1\cdots k)$ and all products of *s*'s obtained by partitioning $1\cdots k$ into subsequences which share the switching point. A negative 1 is associated witch each switching point as is a $e_{HS}$ between the spheres on each side of the switching point.

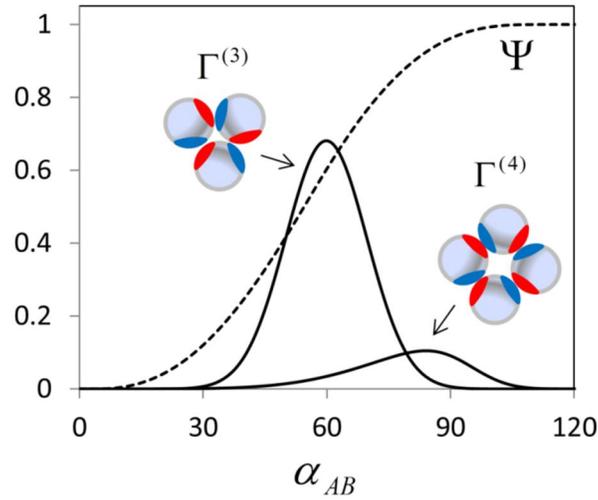

**Figure 3:** Geometric integrals $\Psi$ (dashed curve) and $\Gamma^{(n)}(n=3,4)$ versus bond angle.[7] Integrals were performed using potential parameters $r_c = 1.1d$ and $\theta_c = 27°$

To generalize Eq. (12) to the case of HBC as given through Eqns. (1) – (2) we follow the same logic as in **I**. See **I** for an extensive discussion. Since the first bond in a chain effectively receives an energy $\varepsilon^{(1)}$, and each subsequent bond effectively receives an energy $\varepsilon^{(2)}$, the products of Mayer functions in a chain of length $k$ should be

$$\tilde{f}(1\cdots k) = f_{AB}^{(1)}(12) f_{AB}^{(2)}(23) \cdots f_{AB}^{(2)}(k-1,k) \tag{15}$$

To enforce Eq. (15) in Eq. (12) we simply redefine the functions $s(1\cdots k)$ as

$$s(1\cdots k) \to f_{AB}^{(1)}(12) f_{AB}^{(2)}(23) \cdots f_{AB}^{(2)}(k-1,k) g_{HS}(1\cdots k) \tag{16}$$

The simple transformation introduced in Eq. (16) accounts for the HBC defined by Eqns. (1) – (2). Now the challenge is to evaluate the integrals $E_m$. Since little is known about the correlation functions $g_{HS}(1\cdots k)$ for $k > 2$, we must approximate the higher order $g_{HS}(1\cdots k)$ in superposition. For the current case, a particularly convenient approximation will be the following

$$g_{HS}(1\cdots k) = \prod_{j=1}^{k-1} g_{HS}(r_{j,j+1}) \prod_{i=1}^{k-2} e_{HS}(r_{i,i+2}) \tag{17}$$

The superposition given by Eq. (17) prevents overlap between nearest and next nearest neighbors in the chain and should be most accurate at low densities; it is particularly convenient here, due to the similarity in definition to the $E_m$ integrals. Combining Eqns. (12) – (17) we obtain

$$E_m = \frac{f_{AB}^{(1)} \left(f_{AB}^{(2)} - f_{AB}^{(1)}\right)^{m-1}}{\Omega^m} \int \prod_{k=1}^{m} \left(O_{AB}(k,k+1) g_{HS}(r_{k,k+1})\right) \prod_{j=1}^{m-1} e_{HS}(r_{j,j+2}) d\vec{r}_2 d\Omega_2 \cdots d\vec{r}_{m+1} d\Omega_{m+1} \tag{18}$$

We note that the probability that an isolated associated chain of $m$ bonds and $m + 1$ spheres has a configuration $(123\cdots m+1)$ is given by

$$P_m(123\cdots m+1) = \frac{\prod_{k=1}^{m} O_{AB}(k,k+1)e_{HS}(r_{k,k+1})\prod_{j=1}^{m-1} e_{HS}(r_{j,j+2})}{Z_m} \qquad (19)$$

Where $Z_m$ is the partition function

$$Z_m = \int \prod_{k=1}^{m} O_{AB}(k,k+1)e_{HS}(r_{k,k+1})\prod_{j=1}^{m-1} e_{HS}(r_{j,j+2}) d\vec{r}_2 d\Omega_2 \cdots d\vec{r}_{m+1} d\Omega_{m+1} \qquad (20)$$

Combining Eqns. (18) – (20), and using the definition of the cavity correlation function $y_{HS}(r) = g_{HS}(r)/e_{HS}(r)$ we obtain

$$E_m = f_{AB}^{(1)}\left(f_{AB}^{(2)} - f_{AB}^{(1)}\right)^{m-1} \left\langle \prod_{k=1}^{m} y_{HS}(r_{k,k+1}) \right\rangle \frac{Z_m}{\Omega^m} \qquad (21)$$

where $\langle \ \rangle$ represents an average over the distribution function given by Eq. (19). To an excellent approximation this average can be evaluated as a product of individual averages over the bonding range

$$\left\langle \prod_{k=1}^{m} y_{HS}(r_{k,k+1}) \right\rangle = \langle y_{HS}(r) \rangle_{br}^{m} \qquad (22)$$

where

$$\langle y_{HS}(r) \rangle_{br} = \frac{4\pi \int_d^{r_c} y_{HS}(r) r^2 dr}{4\pi \int_d^{r_c} r^2 dr} = \frac{\xi}{v_b} \qquad (23)$$

The constant $v_b$ is the volume of a spherical shell of thickness $r_c - d$ and $\xi = 4\pi \int_d^{r_c} y_{HS}(r) r^2 dr$. Note $\xi$ is as defined in **II** which differs by a factor of $4\pi$ to the definition in **I**. For $r \geq d$, $y_{HS}(r) = g_{HS}(r)$ which allows Eq. (23) to be easily evaluated using Eq. (8).

Now we can rewrite Eq. (21) as

$$E_m = f_{AB}^{(1)} \left(f_{AB}^{(2)} - f_{AB}^{(1)}\right)^{m-1} \left(\frac{\xi}{v_b}\right)^m \frac{Z_m}{\Omega^m} \tag{24}$$

The partition function $Z_m$ gives the number of associated states a chain consisting of $m + 1$ spheres and $m$ bonds can occupy. To a very good approximation $Z_m$ can be factored as

$$Z_m = Z_1^m \left(\frac{Z_2}{Z_1^2}\right)^{m-1} \tag{25}$$

where we note the definition of $\Psi$ in Eq. (10)

$$\Psi = \frac{Z_2}{Z_1^2} \tag{26}$$

Now combining Eqns. (24) and (25) we obtain the final form for $E_m$

$$E_m = \frac{f_{AB}^{(1)}}{\Psi\left(f_{AB}^{(2)} - f_{AB}^{(1)}\right)} \left(\left(f_{AB}^{(2)} - f_{AB}^{(1)}\right)\Psi\xi\kappa\right)^m \tag{27}$$

where $\kappa = (1 - \cos\theta_c)^2 / 4$. With Eq. (27) we can evaluate the infinite sum in Eq. (11) as

$$\Xi = \sum_{m=1}^{\infty} \rho_o^{m-1} E_m = \frac{f_{AB}^{(1)}\xi\kappa}{1 - \left(f_{AB}^{(2)} - f_{AB}^{(1)}\right)\Psi\xi\kappa\rho_o} \tag{28}$$

In the evaluation of the infinite sum in Eq. (28) we have assumed

$$|\gamma| \leq 1 \quad \text{where} \quad \gamma = \left(f_{AB}^{(2)} - f_{AB}^{(1)}\right)\Psi\xi\kappa\rho_o \tag{29}$$

A similar assumption was made in the corresponding infinite sum in **I** [Eq. (16) of **I**]. We will discuss this point further in section 3. Equation (28) allows for the calculation of $\Delta c_{ch}^{(o)}$ through Eq. (10) as

$$\frac{\Delta c_{ch}^{(o)}}{V} = \frac{\sigma_A \sigma_B f_{AB}^{(1)}\xi\kappa}{1 + \left(f_{AB}^{(1)} - f_{AB}^{(2)}\Psi\right)\xi\kappa\rho_o} \tag{30}$$

Equation (30) completes our analysis for the bond angle dependence of chain formation in two site association fluids with HBC as defined here. It is remarkably simple considering that it accounts for both bond angle dependence and HBC. For large bond angles $\Psi \to 1$ and we recover the result of **I**. In the absence of bond HBC $f_{AB}^{(1)} = f_{AB}^{(2)}$ and we recover the result of **II**.

Now that the Helmholtz free energy has been completely specified we minimize Eq. (4) with respect to $\sigma_B$ and $\rho_o$ to obtain

$$\frac{\sigma_A}{\rho_o} = 1 + \frac{\sigma_A f_{AB}^{(1)} \xi \kappa}{1 + \left(f_{AB}^{(1)} - f_{AB}^{(2)}\Psi\right)\xi \kappa \rho_o} \qquad (31)$$

$$\frac{\rho}{\rho_o} = \left(\frac{\sigma_A}{\rho_o}\right)^2 - \frac{\left(f_{AB}^{(1)} - f_{AB}^{(2)}\Psi\right)}{f_{AB}^{(1)}} \left(\frac{f_{AB}^{(1)} \sigma_A \xi \kappa}{1 + \left(f_{AB}^{(1)} - f_{AB}^{(2)}\Psi\right)\xi \kappa \rho_o}\right)^2 + \sum_{n=3}^{\infty} \frac{\left(f_{AB}^{(2)} \rho_o \hat{g}_{HS} K\right)^n}{\rho_o d^3} \Gamma^{(n)} \qquad (32)$$

In Eq. (32) we have enforced that $\sigma_A = \sigma_B$ due to symmetry. Using Eq. (31) to eliminate $\sigma_A$ in Eq. (32) we have

$$\frac{\rho}{\rho_o} = 1 + \frac{2\rho_o f_{AB}^{(1)} \xi \kappa}{1 - \rho_o \Psi f_{AB}^{(2)} \xi \kappa} + \frac{\left(\rho_o \xi \kappa\right)^2 f_{AB}^{(1)} \Psi f_{AB}^{(2)}}{\left(1 - \rho_o \Psi f_{AB}^{(2)} \xi \kappa\right)^2} + \sum_{n=3}^{\infty} \frac{\left(f_{AB}^{(2)} \rho_o \hat{g}_{HS} K\right)^n}{\rho_o d^3} \Gamma^{(n)} \qquad (33)$$

Equation (33) gives a nonlinear equation for $\rho_o$ which then allows for $\sigma_A$ to be calculated through Eq. (31). Combining the preceding results allows the free energy to be simplified to

$$\frac{A - A_{HS}}{k_B TV} = \rho \ln \frac{\rho_o}{\rho} - \sigma_A + \rho - \sum_{n=3}^{\infty} \frac{\left(f_{AB}^{(2)} \rho_o \hat{g}_{HS} K\right)^n}{nd^3} \Gamma^{(n)} \qquad (34)$$

Equations (31), (33) and (34) give the complete theory for 2 site associating fluids with bond angle dependence and HBC as defined.

For comparison to simulations we will use the fraction of molecules bonded $k$ times which are found as

$$X_o = \rho_o / \rho \qquad X_1 = 2(\sigma_A / \rho - X_o)$$

$$X_2 = 1 - X_o - X_1$$
(35)

To study the distribution of clusters we will need the fraction of spheres in rings of size $n$ which is represented by $\chi_{ring}^{(n)}$ and given by

$$\chi_{ring}^{(n)} = \frac{\left(f_{AB}^{(2)} \rho_o \hat{g}_{HS} K\right)^n}{d^3 \rho} \Gamma^{(n)}$$
(36)

Also, we will use the fraction of spheres bonded at both sites $A$ and $B$ in a linear chain $X_{2c}$ which is found to be

$$X_{2c} = f_{AB}^{(2)} f_{AB}^{(1)} \Psi X_o \left( \frac{\sigma_A \xi \kappa}{1 + \left(f_{AB}^{(1)} - f_{AB}^{(2)} \Psi\right) \xi \kappa \rho_o} \right)^2$$
(37)

From Eq. (37) we see that in the case of total blockage $\Psi \to 0$, or no energetic benefit of forming the second bond $f_{AB}^{(2)} \to 0$, the theory correctly predicts that $X_{2c} \to 0$. Using these defined fractions the free energy Eq. (34) can be rewritten as

$$\frac{A - A_{HS}}{N k_B T} = \ln X_o + 1 - \frac{X_1}{2} - X_o - \sum_{n=3}^{\infty} \frac{\chi_{ring}^{(n)}}{n}$$
(38)

Comparing Eq. (38) to [Eq. (38) of **II**, excluding double bonding and substituting the equality $X_A = X_1/2 + X_o$)] we see the form of the free energy has not changed. In fact, the introduction of bond cooperativity has added negligible additional complexity over the non – cooperative case studied in **II**.

In this section we have derived the first equation of state for associating molecules which explicitly includes the effect of bond angle and HBC. In the next section we compare the new theory to Monte Carlo simulation data.

## 3. Comparison of simulation and theory

To validate the new theory we now compare theoretical predictions to the results of Monte Carlo simulations. We perform all new Monte Carlo simulations in the canonical ensemble using standard[18] methodology. The specific method was outlined in **I**, so for brevity we will not discuss it further here. We choose potential parameters $r_c = 1.1d$ and $\theta_c = 27°$ such that the sites are singly bondable.[19, 20] To isolate our analysis to the effect of bond angle and HBC we perform calculations and simulations at the single moderate liquid like density $\rho^* = \rho d^3 = 0.6$.

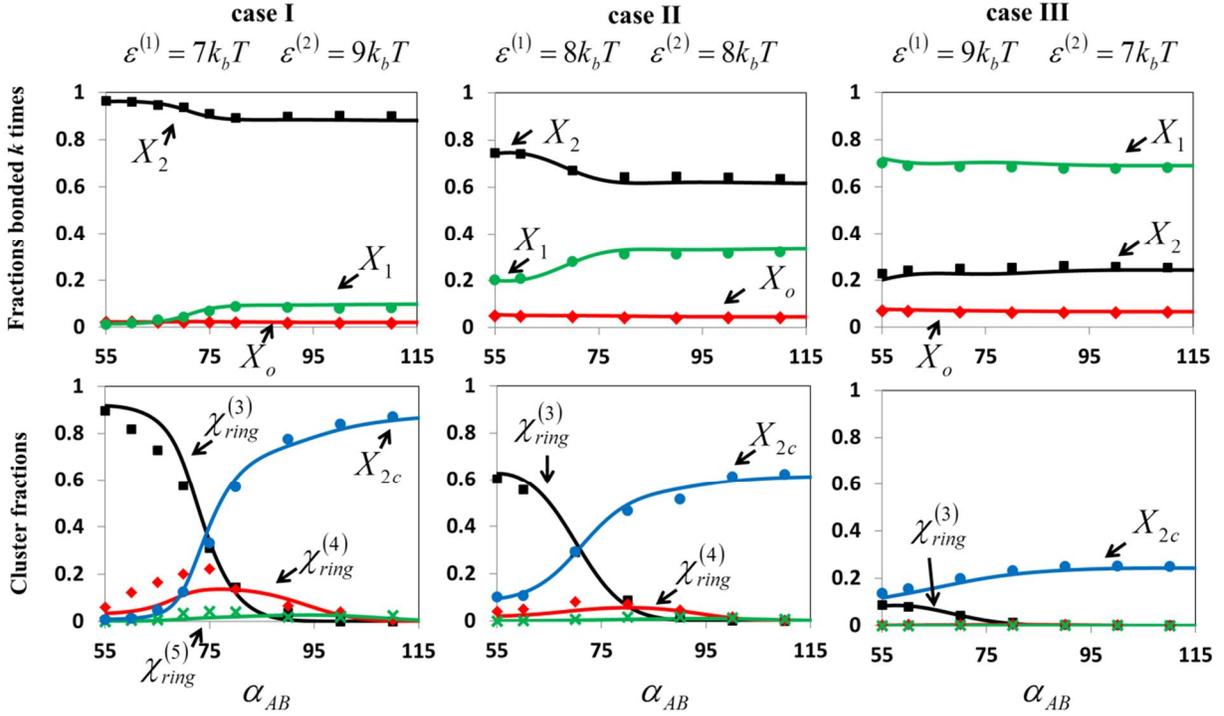

**Figure 4:** Bonding fractions at a density of $\rho^* = 0.6$ for positive HBC (left), no HBC (center) and negative HBC (right). Curves give theoretical predictions and symbols are simulation results. The top row gives the fraction of spheres bonded $k = \{0$ (diamonds), 1 (circles), 2 (squares)$\}$ times. The bottom row gives the fractions bonded twice in a chain $X_{2c}$ (circles) and ring fractions $\chi_{ring}^{(n)}$ $\{n = 3$ (squares), 4 (diamonds) and 5(crosses)$\}$

We will consider 3 cases: in case I $\varepsilon^{(1)} = 7k_BT$ and $\varepsilon^{(2)} = 9k_BT$ meaning bond there is positive HBC, in case II there is no HBC $\varepsilon^{(1)} = 8k_BT$ and $\varepsilon^{(2)} = 8k_BT$, finally in case III there is negative HBC $\varepsilon^{(1)} = 9k_BT$ and $\varepsilon^{(2)} = 7k_BT$. These represent 3 strongly associating systems and will provide a stringent test of the theory.

Figure 4 compares theory and simulation of the bonding fractions $X_k$ (35), $\chi_{ring}^{(n)}$ (36) and $X_{2c}$ (37). We begin our discussion with $X_k$. Comparing the three cases at $\alpha_{AB} = 115°$ we see that $X_2$(case I) $> X_2$(case II) $> X_2$(case III) and $X_1$(case I) $< X_1$(case II) $< X_1$(case III) which is the expected result since case I shows positive HBC and case III shows negative HBC. That is, since $\varepsilon^{(2)} > \varepsilon^{(1)}$ for case I, there is a significant energetic benefit for spheres to become fully bonded, while for $\varepsilon^{(1)} > \varepsilon^{(2)}$, as in case III, it is energetically beneficial to form dimers over longer chains. As the bond angle is decreased from 115° to 55° the monomer fractions $X_o$ remain relatively constant for each case; however there is a much stronger $\alpha_{AB}$ dependence for the fractions $X_2$ and $X_1$. Considering cases I and II, we see that decreasing $\alpha_{AB}$ there is little change in the fractions until $\alpha_{AB} \sim 80°$, while decreasing $\alpha_{AB}$ further results in an increase in $X_2$ and decrease in $X_1$. In case III the $\alpha_{AB}$ dependence of these fractions is much weaker and opposite of the behavior observed in the previous two cases. In case III decreasing $\alpha_{AB}$ results in an increase in $X_1$ and decrease in $X_2$. Comparing theory and simulation for these fractions we see that the theory does an excellent job of predicting the effect of $\alpha_{AB}$ and HBC on the fractions $X_k$.

To explain the behavior of the fractions $X_k$, in Fig. 4 we plot the fractions of spheres bonded twice in the various cluster types. Focusing on cases I and II we see that for $\alpha_{AB} > 105°$ all ring fractions $\chi_{ring}^{(n)}$ are small and the fluid is dominated by chain like clusters ($X_{2c}$ is large). Decreasing $\alpha_{AB}$ below $105°$ steric hindrance begins suppressing chain formation and ring formation becomes more prominent. Over the full range of bond angles, at this current density, only rings of sizes $n = 3 - 5$ exist in significant quantities. The 5 member rings are rare at this density with the 4 member rings becoming more prominent with a maximum $\chi_{ring}^{(4)}$ near $\alpha_{AB} = 80°$. For both cases I and II the triatomic rings become dominant for small bond angles due to the relatively low entropic penalty of association due to the small ring size, and the strong energetic benefit of all spheres in the cluster becoming fully bonded. We also note that for these two cases, the fraction bonded twice in chains $X_{2c}$ becomes small for small $\alpha_{AB}$. The decrease in $X_{2c}$ is the combined effect of steric hindrance between association sites for chain formation, which increases the entropic penalty of association, as well as energetic dominance of the triatomic rings. It is not possible for all spheres to be fully bonded in a chain.

Comparing cases I and II we see that positive HBC (case I) favors ring formation as compared to the non HBC case (case II). This is furthur demonstrated by the fraction $X_{2c}$ which has all but vanished at $\alpha_{AB} = 55°$ for case I while for case II is near $X_{2c} \sim 0.1$. This behavior is a result of the fact that positive HBC favors associated clusters in which all spheres are fully bonded; this can only be realized in ring formation. For this reason, case I shows significantly greater ring formation than case II. For both cases theory is in good agreement with the simulation data.

Now considering case III which shows negative HBC $\varepsilon^{(2)} < \varepsilon^{(1)}$ we see that the ring fractions $\chi_{ring}^{(4)}$ and $\chi_{ring}^{(5)}$ are small over the full bond angle range with the only significant ring contribution coming from $\chi_{ring}^{(3)}$ for $\alpha_{AB} < 80°$. For this case $X_{2c} > \chi_{ring}^{(3)}$ over the full bond angle range, showing that chain formation is always favored. The reason for this is since $\varepsilon^{(2)} < \varepsilon^{(1)}$ much of the energetic benefit of ring formation has been removed. There is an additional entropic penalty for a chain to close and form an associated ring. For case I and II the energetic benefit of ring formation is enough to overcome this penalty; however, for case III this is simply not the case. Again, the theory is in good agreement with the simulation data.

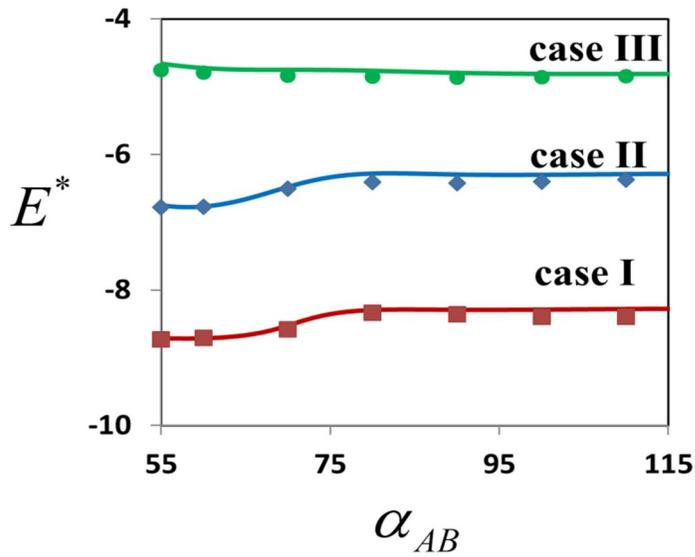

**Figure 5:** Reduced excess internal energy for case I, II and III. Symbols give simulation results and curves give theory predictions.

Now we can see why, for small $\alpha_{AB}$, decreasing $\alpha_{AB}$ results in an increase in $X_2$ and decrease in $X_1$ for cases I and II, while the opposite is true for case III. The reason is ring formation. Decreasing $\alpha_{AB}$ increases the entropic penalty for a sphere to bond twice in a chain. When ring formation is small, this necessarily results in an increase in $X_1$ and decrease in $X_2$, as in case III; however, when ring formation is favored, decreasing the bond angle results in an increase in ring formation which overcomes this troublesome association into chains. This results in an increase in $X_2$ and decrease in $X_1$. Looking back at Fig. 4, it is remarkable that the simple theory derived in this paper accurately accounts for this complex behavior.

Figure 5 compares the excess internal energy for these three cases. As noted above, association is strongest in case I and weakest in case III which gives the following relation among the internal energies $E$ (case I) $< E$ (case II) $< E$ (case III). For cases I and II decreasing $\alpha_{AB}$ below $\alpha_{AB} \sim 80°$ results in a decrease in $E$, while for case III decreasing $\alpha_{AB}$ below $\alpha_{AB} \sim 80°$ results in an increase in $E$. Like the trends noted for $X_2$, the bond angle dependence of $E$ can be traced back to ring formation. For each case decreasing $\alpha_{AB}$ necessarily inhibits chain formation due to steric hindrance. For cases I and II additional ring formation at small bond angles more than makes up for the decrease in chain formation and results in a decrease in $E$. For case III ring formation is much smaller, not enough to make up for the decrease in association into chains at small $\alpha_{AB}$, which results in an increase in $E$ as $\alpha_{AB}$ is decreased. Theory and simulation are in excellent agreement.

We conclude this section with a brief discussion of the condition given by Eq. (29) that $|\gamma| \leq 1$ for the evaluation of the infinite sum given by Eq. (28). For cases I and II, with positive and no HBC respectively, this condition is easily satisfied at each bond angle. However, for case III, with negative HBC, $1.61 < \gamma < 2.72$ which is in clear violation of Eq. (29). Also, analyzing the results presented in **I** we find that $|\gamma| \leq 1$ for all cases of positive HBC $\varepsilon^{(2)} > \varepsilon^{(1)}$ with the only instances of $|\gamma| > 1$ occurring for the case of strong negative HBC with $\varepsilon^{(2)} < \varepsilon^{(1)} - k_B T$. Of course, both the results of this paper and **I** show that the theory is in excellent agreement with simulation for cases which exhibit strong negative HBC with $|\gamma| > 1$. This shows that the final equations derived assuming $|\gamma| \leq 1$ are also accurate for $|\gamma| > 1$. Furthermore, in nature HBC

arises from the fact that when a multi – functional hydrogen bonding molecule forms multiple hydrogen bonds the polarization of the molecule is increased.[21] This necessarily results in positive HBC only, for which the condition $|\gamma| \leq 1$ seems to always hold.

## 4. Conclusions

We have developed a new equation of state for associating fluids with two association sites. Using resummed perturbation theory, we have included both the effects of bond angle and HBC for the first time. The resulting equation of state is surprisingly simple with negligible additional complexity over the non – HBC[7] case. It was shown that both bond angle and HBC play a huge role in the types of associated clusters which are formed. In agreement with detailed quantum calculations[9], we have shown that positive HBC favors ring formation. To test the theory new Monte Carlo simulations were performed. The theory was found to be accurate.

## Acknowledgments

The financial support of The Robert A. Welch Foundation Grant No. C – 1241 is gratefully acknowledged

## Appendix:

In this appendix we derive the chemical potential $\mu$, excess internal energy $E$ and pressure $P$ from the results of section 2. The chemical potential is obtained from the general relation

$$\frac{\mu}{k_B T} = \frac{\mu_{HS}}{k_B T} + \ln \frac{\rho_o}{\rho} - \frac{\partial \Delta c^{(o)}/V}{\partial \rho} \tag{A1}$$

We calculate the derivative as

$$\frac{\partial \Delta c^{(o)}/V}{\partial \rho} = \sum_{n=3}^{\infty} \frac{\left(f_{AB}^{(2)} \rho_o \hat{g}_{HS} K\right)^n}{d^3} \Gamma^{(n)} \frac{\partial \ln \hat{g}_{HS}}{\partial \rho} + \frac{\sigma_A^2 f_{AB}^{(1)} \xi \kappa}{1 + \left(f_{AB}^{(1)} - f_{AB}^{(2)} \Psi\right) \xi \kappa \rho_o} \frac{\partial \ln \xi}{\partial \rho} \quad (A2)$$
$$- \frac{\left(f_{AB}^{(1)} - f_{AB}^{(2)} \Psi\right)}{f_{AB}^{(1)}} \rho_o \left(\frac{\sigma_A f_{AB}^{(1)} \xi \kappa}{1 + \left(f_{AB}^{(1)} - f_{AB}^{(2)} \Psi\right) \xi \kappa \rho_o}\right)^2 \frac{\partial \ln \xi}{\partial \rho}$$

Which can be simplified to

$$\frac{\partial \Delta c^{(o)}/V}{\partial \rho} = \rho \frac{\partial \ln \hat{g}_{HS}}{\partial \rho} \sum_{n=3}^{\infty} \chi_{ring}^{(n)} + \left(\frac{\sigma_A}{\rho_o} - 1\right)\left(\sigma_A - \frac{\left(f_{AB}^{(1)} - f_{AB}^{(2)}\Psi\right)}{f_{AB}^{(1)}} \rho_o \left(\frac{\sigma_A}{\rho_o} - 1\right)\right) \frac{\partial \ln \xi}{\partial \rho} \quad (A3)$$

With the chemical potential known the pressure is easily calculated through the relation

$$P = \mu \rho - A/V \quad (A4)$$

Now all that remains is the calculation of the excess internal energy which is given by

$$\frac{E}{N} = \frac{\partial}{\partial \beta} \frac{\beta A}{N} = \frac{\dot{\rho}_o}{\rho_o} - \frac{\dot{\sigma}_A}{\rho} - \left(\frac{\dot{\rho}_o}{\rho_o} + \frac{\dot{f}_{AB}^{(2)}}{f_{AB}^{(2)}}\right) \sum_{n=3}^{\infty} \chi_{ring}^{(n)} \quad (A5)$$

In Eq. (A5) $\beta = 1/k_B T$ and $\dot{a} = \partial a / \partial \beta$. Taking the derivative of Eq. (32) we obtain

$$-\frac{\rho \dot{\rho}_o}{\rho_o^2} = 2\rho_o \frac{f_{AB}^{(1)} \xi \kappa}{1 - \rho_o \Psi f_{AB}^{(2)} \xi \kappa} \left(\frac{\dot{\rho}_o}{\rho_o} + \frac{\dot{f}_{AB}^{(1)}}{f_{AB}^{(1)}}\right) + 2\rho_o^2 \frac{f_{AB}^{(2)}}{f_{AB}^{(1)}} \Psi \left(\frac{f_{AB}^{(1)} \xi \kappa}{1 - \rho_o \Psi f_{AB}^{(2)} \xi \kappa}\right)^2 \left(\frac{\dot{\rho}_o}{\rho_o} + \frac{\dot{f}_{AB}^{(2)}}{f_{AB}^{(2)}}\right) \quad (A6)$$
$$+ \rho_o^2 \frac{f_{AB}^{(2)}}{f_{AB}^{(1)}} \Psi \frac{\left(\xi \kappa f_{AB}^{(1)}\right)^2}{\left(1 - \rho_o \Psi f_{AB}^{(2)} \xi \kappa\right)^2} \left(2\frac{\dot{\rho}_o}{\rho_o} + \frac{\dot{f}_{AB}^{(1)}}{f_{AB}^{(1)}} + \frac{\dot{f}_{AB}^{(2)}}{f_{AB}^{(2)}}\right) + 2\rho_o^3 \left(\Psi \frac{f_{AB}^{(2)}}{f_{AB}^{(1)}}\right)^2 \left(\frac{\xi \kappa f_{AB}^{(1)}}{1 - \rho_o \Psi f_{AB}^{(2)} \xi \kappa}\right)^3 \left(\frac{\dot{\rho}_o}{\rho_o} + \frac{\dot{f}_{AB}^{(2)}}{f_{AB}^{(2)}}\right)$$
$$+ \sum_{n=3}^{\infty} \frac{\left(f_{AB}^{(2)} \rho_o \hat{g}_{HS} K\right)^n}{\rho_o d^3} \Gamma^{(n)} \left((n-1)\frac{\dot{\rho}_o}{\rho_o} + n \frac{\dot{f}_{AB}^{(2)}}{f_{AB}^{(2)}}\right)$$

and solving for $\dot{\rho}_o$

$$\frac{\dot{\rho}_o}{\rho_o} = -\frac{\left(2\delta + \gamma \delta^2\right)\frac{\dot{f}_{AB}^{(1)}}{f_{AB}^{(1)}} + \left(2\gamma^2 \delta^3 + 3\gamma \delta^2 + \frac{\rho}{\rho_o} \sum_{n=3}^{\infty} n \chi_{ring}^{(n)}\right) \frac{\dot{f}_{AB}^{(2)}}{f_{AB}^{(2)}}}{\frac{\rho}{\rho_o} + 2\delta + 4\delta^2 \gamma + 2\delta^3 \gamma^2 + \frac{\rho}{\rho_o} \sum_{n=3}^{\infty} (n-1)\chi_{ring}^{(n)}} \quad (A7)$$

Where we have defined

$$\delta = \frac{\rho_o f_{AB}^{(1)} \xi \kappa}{1 - \rho_o \Psi f_{AB}^{(2)} \xi \kappa} \tag{A8}$$

$$\gamma = \frac{f_{AB}^{(2)}}{f_{AB}^{(1)}} \Psi$$

Lastly $\dot{\sigma}_A$ is obtained from Eq. (30) as

$$\dot{\sigma}_A = \dot{\rho}_o + (\sigma_A - \rho_o)\left(\frac{\dot{\rho}_o}{\rho_o} + \frac{\dot{\sigma}_A}{\sigma_A} + \frac{\dot{f}_{AB}^{(1)}}{f_{AB}^{(1)}}\right) - \frac{1}{\sigma_A}(\sigma_A - \rho_o)^2 \left(\frac{\left(\dot{f}_{AB}^{(1)} - \dot{f}_{AB}^{(2)}\Psi\right)}{f_{AB}^{(1)}} + \frac{\left(f_{AB}^{(1)} - f_{AB}^{(2)}\Psi\right)}{f_{AB}^{(1)}} \frac{\dot{\rho}_o}{\rho_o}\right) \tag{A9}$$

Solving of for $\dot{\sigma}_A$ we obtain

$$\dot{\sigma}_A = \frac{\dot{\rho}_o + (\sigma_A - \rho_o)\left(\frac{\dot{\rho}_o}{\rho_o} + \frac{\dot{f}_{AB}^{(1)}}{f_{AB}^{(1)}}\right) - \frac{1}{\sigma_A}(\sigma_A - \rho_o)^2 \left(\frac{\left(\dot{f}_{AB}^{(1)} - \dot{f}_{AB}^{(2)}\Psi\right)}{f_{AB}^{(1)}} + \frac{\left(f_{AB}^{(1)} - f_{AB}^{(2)}\Psi\right)}{f_{AB}^{(1)}} \frac{\dot{\rho}_o}{\rho_o}\right)}{1 - \frac{\sigma_A - \rho_o}{\sigma_A}} \tag{A10}$$

Equations (A5), (A7) and (A10) give the internal energy.

## References:

1. M. Wertheim, Journal of Statistical Physics **35** (1), 19-34 (1984).
2. M. Wertheim, Journal of Statistical Physics **35** (1), 35-47 (1984).
3. M. Wertheim, Journal of Statistical Physics **42** (3), 459-476 (1986).
4. M. Wertheim, Journal of Statistical Physics **42** (3), 477-492 (1986).
5. M. Wertheim, The Journal of Chemical Physics **87**, 7323 (1987).
6. W. G. Chapman, G. Jackson and K. E. Gubbins, Molecular Physics **65** (5), 1057-1079 (1988).
7. B. D. Marshall and W. G. Chapman, Physical Review E **87** (5), 052307 (2013).
8. B. D. Marshall and W. G. Chapman, The Journal of chemical physics **139** (5), 054902 (2013).
9. L. Rincón, R. Almeida, D. García-Aldea and H. D. y Riega, The Journal of Chemical Physics **114**, 5552 (2001).

10. S. J. Barlow, G. V. Bondarenko, Y. E. Gorbaty, T. Yamaguchi and M. Poliakoff, The Journal of Physical Chemistry A **106** (43), 10452-10460 (2002).
11. H. Guo and M. Karplus, The Journal of Physical Chemistry **98** (29), 7104-7105 (1994).
12. B. D. Marshall and W. G. Chapman, The Journal of chemical physics **139** (21), 214106 (2013).
13. R. P. Sear and G. Jackson, The Journal of Chemical Physics **105** (3), 1113 (1996).
14. W. Bol, Molecular Physics **45**, 605 (1982).
15. G. Jackson, W. G. Chapman and K. E. Gubbins, Molecular Physics **65** (1), 1-31 (1988).
16. N. Kern and D. Frenkel, The Journal of Chemical Physics **118**, 9882 (2003).
17. W. G. Chapman, PhD. Thesis. 1988, Cornell University: Ithaca, NY.
18. D. Frenkel and B. Smit, *Understanding molecular simulation: from algorithms to applications*. (Academic press, 2001).
19. Y. Kalyuzhnyi, H. Docherty and P. Cummings, The Journal of Chemical Physics **133**, 044502 (2010).
20. B. D. Marshall, D. Ballal and W. G. Chapman, The Journal of Chemical Physics **137** (10), 104909 (2012).
21. A. E. Reed, L. A. Curtiss and F. Weinhold, Chemical Reviews **88** (6), 899-926 (1988).